\documentclass[sn-mathphys-num]{sn-jnl}

\usepackage{graphicx}%
\usepackage{multirow}%
\usepackage{amsmath,amssymb,amsfonts}%
\usepackage{amsthm}%
\usepackage{mathrsfs}%
\usepackage[title]{appendix}%
\usepackage{xcolor}%
\usepackage{textcomp}%
\usepackage{manyfoot}%
\usepackage{booktabs}%
\usepackage{algorithm}%
\usepackage{algorithmicx}%
\usepackage{algpseudocode}%
\usepackage{listings}%
\usepackage{bm}%
\usepackage{geometry}
\theoremstyle{thmstyleone}%
\theoremstyle{thmstyletwo}%

\theoremstyle{thmstylethree}%

\begin{document}

\title[pADM]{Adomian decomposition method reformulated using dimensionless nonlinear perturbation theory}


\author{\fnm{Albert S.} \sur{Kim}}\email{albertsk@hawaii.edu}
\affil{\orgdiv{Civil, Environmental and Construction Engineering}, \orgname{University of Hawai'i at Manoa}, \orgaddress{\street{2540 Dole Street, Holmes 383}, \city{Honolulu}, \postcode{96822}, \state{HI}, \country{USA}}}






\abstract{The Adomian decomposition method (ADM) is a universal approach to
solving governing equations in various engineering and technological
applications.  The applicability of the ADM is almost limitless due
to its universal applicability, but its convergence rate and numerical
accuracy are sensitive to the number of truncated terms in series
solutions. More importantly, Adomian formalism still holds unresolved
issue regarding the mismatch of the order of the expansion parameter.
The current work provides an in-depth analysis of Adomian's decomposition
method, Lyapunov\textquoteright s stability theory, and the nonlinear
perturbation theory to resolve the fundamental mismatch with physical
interpretation.}




\maketitle

\section{Introduction}

The Adomian Decomposition Method (ADM) is a mathematical technique
that uses iterative operator theory to solve nonlinear differential
equations, including stochastic differential equations 
\citep{Adomian1963,Adomian1976,Adomian.Malakian1980,Adomian.Malakian1980a,Adomian1983,Adomian.Bellomo.ea1983}.
One of the notable benefits of the ADM is its ability to be applied
universally to a wide range of governing equations, such as ordinary
and partial differential equations and multi-dimensional integrals
in various physical and engineered systems \citep{Adomian1986,Adomian1990,Adomian1991,adomianSolvingFrontierProblems1994}.
Advanced ADM approaches were applied in various areas of applied mathematics
and physics, as follows. The variational iteration method was developed
to solve boundary value problems on finite and semi-infinite intervals,
enabling the easy identification of the Lagrange multiplier and providing
a computational advantage for converging approximate solutions on
semi-infinite intervals \citep{Mohyud-DinEtAl2017}. In the dynamics
of hydromagnetic ferro-nanofluid, the governing equations are transformed
into ordinary differential equations (ODEs) and solved analytically
using the ADM that was compared with numerical solutions obtained
using the Runge-Kutta-Fehlberg method \citep{HamrelaineEtAl2022}.
A time fractional partial integro-differential equation was solved
using the ADM and compared with the homotopy perturbation method,
providing an in-depth analysis of the convergence to the exact solution
\citep{PandaEtAl2022}. To overcome singularity at the origin, multi-point
singular boundary value problems are converted into equivalent integral
equations, and the ADM is applied with convergence analysis for approximate
solutions \citep{SinghWazwaz2022}. A nonlinear partial differential
equation in 4D Euclidean anti-de Sitter space was solved, providing
perturbative series solutions for three specific modes of the equation,
all valid in the probe approximation \citep{Naghdi2023}.

The ADM\textquoteright s application scopes and disciplinary areas
can be further expanded by resolving its intrinsic limitations, as
follows. First, the convergence rate of ADM\textquoteright s series
solutions is impacted by the nonlinear nature of the governing equation.
With increasing $n$, higher-order terms receive more low-order coefficients
\citep{Abdelrazec.Pelinovsky2011}. A new analytic method was proposed
\citep{Zhang.Zou.ea2018} which includes a convergence-control parameter
to regulate the convergence region and rate; however, an additional
numerical algorithm is required to identify the optimal parameter
value. Second, the ADM fails to provide reliable solutions for initial
value problems at asymptotic limits \citep{Abdelrazec.Pelinovsky2011},
likely due to the limited availability of the maximum ADM polynomial
($n\le10$) \citep{Adomian1991,adomianSolvingFrontierProblems1994,Adomian1995}.
The construction of extended tables becomes progressively more demanding
in terms of effort and complexity as the order $n$ is raised. Nevertheless,
there have been additional methodologies that prioritize expediting
evaluations of $A_{n}$ (rather than incorporating higher order terms
based on Adomian's original research), including modified \citep{Adomian.Rach1996,Al-Shareef.AlQarni.ea2016,Kumar.Umesh2022},
iterative \citep{BiazarEtAl2010}, and multi-term \citep{Daftardar-GejjiBhalekar2008,SadeghiniaKumar2021}
ADMs.  Third, the ADM is facing an unresolved issue concerning the
inconsistency of the expansion parameter $\lambda$ between the polynomial
coefficient and the recurrence formula \citep{Zhang.Liang2015}. The
main focus should be on meticulously resolving the Adomian's dilemma,
the discrepancy issue addressed in this work. 

The terms constituting $A_{n}$ were obtained by Adomian and Rach
\citep{Adomian.Rach1983,Rach1984,Rach2008} through the utilization
of a series parameter, reminiscent of the method employed in conventional
perturbation theory: nevertheless, they underscored the autonomy of
the ADM from perturbation and analogous approaches. Nevertheless,
the series expansions of the general nonlinear governing equation
were not correctly segregated and synchronized with respect to the
parameter order. The analysis of the available literature accentuates
the divergence in how nonlinear perturbation theory and Lyapunov's
stability theory might discern the mathematical formalism of ADM,
potentially yielding physical interpretations when applied to physical
and engineering systems. Dynamical systems in diverse fields are assessed
for stability through the application of the LST. This analysis is
crucial in determining the stability of a system's equilibrium point,
categorizing it as stable, asymptotically stable, or unstable. Multiple
mathematicians and physicists have developed perturbation theory over
time \citep{Bell1986}, including Lagrange \citep{Lagrange1772},
Laplace \citep{Laplace1798}, and Poincare \citep{Poincare1890} in
celestial dynamics and Schrodinger \citep{Schrodinger1926,Schrodinger1926a},
Dirac \citep{Dirac1955}, and Feynman \citep{Feynman1949} in quantum
mechanics.

The resolution of the Adomian's dilemma is of paramount importance
in rigorously improving the quality of the ADM solution in diverse
areas. The present work provides a comprehensive summary of ADM formalism,
followed by recent endeavors to tackle the parameter order mismatch
issue. Subsequently, a theoretical resolution is presented to address
the parameter order mismatch through the fundamental perturbation
formalism.

\section{Problem Statement}

The operator equation in the Adomian literature can be written as
\begin{equation}
L\left[U\right]+R\left[U\right]+N\left[U\right]=g\label{eq:ADM-goveq}
\end{equation}
where $L$, $R$, and $N$ represent linear, remaining, and nonlinear
operators, respectively, for a specific physical quantity $U$. In
this work, time $t$ is selected as an independent variable for an
$s$-fold differential operator $L$, such as $L\left[\cdots\right]=\partial^{s}\left[\cdots\right]/\partial t^{s}$,
which is assumed to be invertible, such as 
\begin{equation}
L^{-1}\left[\cdots\right]=\int^{t}dt_{1}\cdot\int^{t_{1}}dt_{2}\cdots\int^{t_{s-1}}dt_{s}\left[\cdots\right]
\end{equation}
The remaining linear operator $R$ includes lower-order derivatives,
such as $\partial^{k}\left[\cdots\right]/\partial t^{k}$ for $k$
from 0 to $s-1$ and also $r$-th order spatial derivatives $\partial^{r}\left[\cdots\right]/\partial q_{k}^{r}$
for the $k$-th generalized coordinate $q_{k}\,\left(k=1-d\right)$
in a $d$-dimensions. For example, if a 3D thermal diffusion is involved,
$L$ contains a first-order time differential, and $R$ includes a
Laplacian operator (of $r=2$) with generalized coordinates of $\bm{q}=\left(q_{1},q_{2},q_{3}\right)=\left(x,y,z\right)$,
where $U$ is temperature $T$ {[}K{]}. The target variable $U$,
especially for transport phenomena, can be also concentration $C$
{[}mol/liter{]} in the molecular diffusion equations, fluid velocity
vector $\bm{V}$ {[}m/s{]} in the Navier-Stokes equation, and probability
distribution functions in the Fokker-Planck and Schrodinger equations.
Throughout the present work, $O\left[U\right]$ means operator $O$
operated to variable $U$, $g\left(t\right)$ specifies $g$ as a
function of $t$, and $f\left(\left\{ x_{n}\right\} \right)$ indicates
$f\left(x_{0},x_{1},\cdots,x_{n}\right)$, i.e., a function of $n+1$
independent variables from $x_{0}$ to $x_{n}$.

Adomian suggested a decomposed series of $U$ \citep{Adomian1991},
such as 
\begin{equation}
U=U_{0}+U_{1}+\cdots=\sum_{n=0}^{\infty}U_{n}\label{eq:U-series}
\end{equation}
where $U_{n}$ is denoted as the $n$-th component function of $U$.
An identity operation was employed such as $L^{-1}\left[L\left[U\right]\right]=U-U\left(0\right)$,
where $U\left(0\right)$ is a complementary solution of $s$-fold
Eq. \eqref{eq:ADM-goveq}, such as 
\begin{equation}
U\left(0\right)=\sum_{k=0}^{s-1}\frac{t^{k}}{k!}\left[\frac{\partial^{k}U}{\partial t^{k}}\right]_{t=0}
\end{equation}
and the inverse operator $L^{-1}$ was applied on the both sides of
Eq. \eqref{eq:ADM-goveq} from the left to result in 
\begin{align}
U-U_{0} & =-L^{-1}\left[R\left[U\right]+N\left[U\right]\right]\label{eq:U-evol}
\end{align}
where $U_{0}=U\left(0\right)+L^{-1}g$ is a trivial solution for $R=N=0$.
Eq. \eqref{eq:U-series} was substituted into Eq. \eqref{eq:U-evol}
to match the terms having the same appearing sequences on each side.
Then, the recurrence formula was obtained, such as 
\begin{align}
U_{n+1} & =-L^{-1}R\left[U_{n}\right]-L^{-1}A_{n}\left(\left\{ U_{n}\right\} \right)\label{eq:ADM-recur}
\end{align}
where $A_{n}\left(\left\{ U_{n}\right\} \right)$ is the Adomian polynomial,
determined by letting $N\left[U\right]$ as a continuous function,
expanded using Taylor's series, such as 
\begin{equation}
N\left[U\right]=f\left(U\right)=\sum_{n=0}\frac{f_{0}^{\left(n\right)}}{n!}\left(U-U_{0}\right)^{n}=\sum_{n=0}^{\infty}A_{n}\left(\left\{ U_{n}\right\} \right)
\end{equation}
where $f_{0}^{\left(n\right)}=\left[d^{n}f\left(U\right)/dU^{n}\right]_{U=U_{0}}$
is the $n^{\mathrm{th}}$ derivative of $f\left(U\right)$ evaluated
at $U_{0}$. To derive the series solution $U$, a generating function
of $A_{n}$ is crucial (instead of its manual calculation for specific
$n$) derived in principle, such as
\begin{equation}
A_{n}=\frac{1}{n!}\left[\frac{\partial^{n}}{\partial\lambda^{n}}f\left(U\left(\lambda\right)\right)\right]_{\lambda=0}=\sum_{k=0}^{n}f_{0}^{\left(k\right)}\,C_{k,n}\label{eq:An-w-lambda}
\end{equation}
where $C_{k,n}$ are a product (or a sum of products) of $k$ component
functions, whose subscripts sum to $n$, divided by the factorial
of the repeated number of subscripts \citep{Adomian.Rach1983,Rach1984,Rach2008}
(See Appendix for an exemplary description), and $U\left(\lambda\right)$
is a perturbation-style series as an alternative representation of
Eq. \eqref{eq:U-series}, such as
\begin{equation}
U\left(\lambda\right)=U_{0}+\lambda U_{1}+\lambda^{2}U_{2}+\cdots=\sum_{n=0}^{\infty}\lambda^{n}U_{n}\label{eq:U-lambda}
\end{equation}
where $\lambda$ is often defined as an expansion parameter. Adomian's
original work independently use Eq. \eqref{eq:U-series} and \eqref{eq:U-lambda}
for separate purposes, as follows. The decomposed series $U$ of Eq.
\eqref{eq:U-series} was directly substituted into Eq. \eqref{eq:ADM-goveq}
(without using $\lambda$) to create the recurrence formula of Eq.
\eqref{eq:ADM-recur}; but, the perturbation-style series $U$ of
Eq. \eqref{eq:U-lambda} was employed only to collect terms for $A_{n}$
comprising $U_{0}$, $U_{1}$, ..., and $U_{n}$ using $\lambda$.
However, direct substituting Eq. \eqref{eq:U-lambda} into \eqref{eq:ADM-goveq}
does not reproduce the same recurrence formula of Eq. \eqref{eq:ADM-recur},
which is denoted as Adomian's dilemma in this work.

Related topics to the efficient expansion were discussed to provide
more robust representations and efficient computing algorithms for
Adomian polynomials \citep{Wazwaz2000,Fatoorehchi.Abolghasemi2016,katariaSimpleParametrizationMethods2016}.
In a new algorithm \citep{Wazwaz2000}, $U$ was used as a sum of
its components, $U_{n}$ for $n\ge1$ (instead of those for $n\ge0$),
which can be elegantly used without any extra formulas other than
elementary operations. The orthogonality of complex variable $e^{inx}$
for integer $n$ was introduced to compute  Adomian polynomials for
various nonlinear operators \citep{katariaSimpleParametrizationMethods2016}
without continuously storing the index sum of $U_{n}$ \citep{Wazwaz2000}.
Alternatively the final solution $U$ (to be sought) is pre-expanded
as a Taylor series of the independent variable (such as $t$, used
in this study) and substituted into the nonlinear term of $N\left[U\right]$,
which reduces computational costs for mere integration and simple
arithmetic operations \citep{Fatoorehchi.Abolghasemi2016,fatoorehchiImprovingDifferentialTransform2013}.
The above-mentioned approaches \citep{Wazwaz2000,fatoorehchiImprovingDifferentialTransform2013,Fatoorehchi.Abolghasemi2016,katariaSimpleParametrizationMethods2016}
provided new insight and faster algorithms to solve nonlinear differential
equations. However, the order of expansion parameter $\lambda$ remained
mismatched, providing series solutions at the same level of accuracy
and convergence as Adomian's original work.

The parameter order mismatch has not been theoretically discussed
until Zhang and Liang \citep{Zhang.Liang2015} initiated the fundamental
issue. Rach \citep{Rach1984} emphasized that ``$\lambda$ is not
in any sense a perturbation parameter but a convenient device for
collecting terms and is dropped at the end of the calculation.''
In fact, Adomian and Rach \citep{Adomian.Rach1983} introduced $\lambda$
as an intermediate parameter to collect terms satisfying conditions
in Eq. \eqref{eq:An-w-lambda}. At the end, $\lambda$ was replaced
by 0 to nullify terms having higher orders than $n$ because terms
of lower orders than $n$ were already eliminated by $n$-time differentiation.
Adomian's explicit expression $A_{n}$ of Eq. \eqref{eq:An-w-lambda}
originates from the perturbation-style series of Eq. \eqref{eq:U-lambda}.
In this case, the nonlinear term can be rigorously rewritten, such
as
\begin{equation}
N\left[U\right]=\lim_{\lambda\to1}\,\sum_{n=0}^{\infty}\lambda^{n}A_{n}\left(\left\{ U_{n}\right\} \right)=\lim_{\lambda\to1}\,\sum_{n=0}^{\infty}\lambda^{n}\frac{1}{n!}\left[\frac{\partial^{n}}{\partial\lambda^{n}}f\left(U\left(\lambda\right)\right)\right]_{\lambda=0}
\end{equation}
where $\lambda^{n}$ inside the summation will be replaced by 1 after
$A_{n}$ is obtained. Zhang and Liang \citep{Zhang.Liang2015} multiplied
the small parameter $\lambda$ to the remainder and nonlinear operator
terms, such as
\begin{equation}
L\left[U\right]+\lambda\left(R\left[U\right]+N\left[U\right]\right)=g\label{eq:ZhangLiang}
\end{equation}
and indicated that the ADM must be a special case of LASPM, which
seemed successfully resolving the parameter order mismatch by providing
some clues to the ``mysterious relationship.'' The adjusted Eq.
\eqref{eq:ZhangLiang} differs from physical governing equations that
do not explicitly contain $\lambda$. Introducing the additional $\lambda$
in Eq. \eqref{eq:ZhangLiang} is at the same level of Adomian's intuitive
ordering used in Eq. \eqref{eq:U-evol}, and  the source-term $g$
remained unmodified.

Lyapunov\textquoteright s theory, primarily assessing system stability
\citep{Kao.Jiang2005}, uses an artificially small parameter, known
as $\varepsilon$, having a specific dimension that can be determined
empirically \citep{Benettin.Galgani.ea1980,Wolf.Swift.ea1985}. Adomian's
expansion parameter, $\lambda$, was used as a temporary derivation
tool to derive the analytical expression of $A_{n}$ and the recurrence
Eq. \eqref{eq:ADM-recur} without explicitly including $\lambda$
at the end. Therefore, it is challenging to fit one theory into the
framework of another, and a detailed physical interpretation is important
for achieving more rigorous resolution. We used the standard perturbation
theory to provide a physics-based resolution of Adomian's dilemma.

\section{Perturbation-based analysis of Adomian polynomial}

We reformulate Adomian's governing equation in the perturbation-style
formalism and provide a physics-based resolution of the Adomian's
dilemma.

\subsection{Nondimensionalization of Adomian Polynomial}

Adomian considered the nonlinear term $N\left[U\right]$ as a Taylor
series of $U$ evaluated at $U_{0}$, where $U_{0}$ is a trivial
solution in the absence of $R\left[U\right]$ and $N\left[U\right]$.
In transport phenomena, nonlinear terms are often found as products
of $U$ and its derivatives, such as, $U\frac{\partial U}{\partial t}$,
$U\frac{\partial U}{\partial x}$, and $U\frac{\partial^{2}U}{\partial x^{2}}$,
which require to extend $N\left[U\right]$ to $N\left[U,\dot{U},U^{\prime}\right]=f\left(U,\dot{U},U^{\prime}\right)$,
where $\dot{U}=\partial U/\partial t$ and $U^{\prime}=\partial U/\partial x$.
The coupling between $U$, $\dot{U}$, and $U^{\prime}$ can be represented
by including three parts separately, such as 
\begin{align}
f\left(U\left(\lambda\right),\dot{U}\left(\lambda\right),U^{\prime}\left(\lambda\right)\right) & =\sum_{n=0}^{\infty}\lambda^{n}A_{n}\left(U_{0},\cdots,U_{p},\dot{U}_{0},\cdots,\dot{U}_{q},U_{0}^{\prime},\cdots,U_{r}^{\prime}\right)
\end{align}
where $p+q+r=n$. For dimensionless analysis, one can select representative
parameters of $U$, $\dot{U}$, and $U^{\prime}$, denoted as $\overline{U}$,
$\overline{V}$, and $\overline{U^{\prime}}$, respectively, to propose
dimensionless variables, such as $u=U/\overline{U}$, $\dot{u}=\dot{U}/\overline{V}$,
and $u^{\prime}=U^{\prime}/\overline{U^{\prime}}$. Because $A_{n}$
is a sum of component products, the scalability of $A_{n}$ is straightforward,
such as 
\begin{equation}
A_{n}\left(U_{0},\cdots,U_{p},\dot{U}_{0},\cdots,\dot{U}_{q},U_{0}^{\prime},\cdots,U_{r}^{\prime}\right)=\alpha_{n}^{*}A_{n}\left(u_{0},\cdots,u_{p},\dot{u}_{0},\cdots,\dot{u}_{q},u_{0}^{\prime},\cdots,u_{r}^{\prime}\right)
\end{equation}
where $\alpha_{n}^{*}=\left(\overline{U}\right)^{p}\left(\overline{V}\right)^{q}\left(\overline{U^{\prime}}\right)^{r}$
is a scaling parameter of $A_{n}$ being a function of dimensionless
variables. Including higher orders of time and spatial derivatives
in the nonlinear term does not alter the scaling property of $A_{n}$.

\subsection{Ordering mismatch resolved}

Partial differential equations often used in the transport and wave
phenomena can be generally written as 
\begin{equation}
\sum_{k=1}^{s}a_{k}D_{t}^{k}U=\sum_{m=1}^{r}\sum_{i=1}^{d}b_{m}\mathscr{D}_{q_{i}}^{m}U+N\left(U,\left\{ \mathscr{D}_{q_{i}}^{m}U\right\} ,\left\{ D_{t}^{s-1}U\right\} \right)+g\left(t,\left\{ q_{i}\right\} \right)\label{eq:trans-eq}
\end{equation}
where $D_{t}^{n}U$ is the partial or ordinary time differentials,
such as $\partial^{n}U/\partial t^{n}$ or $d^{n}U/dt^{n}$, respectively,
$\mathscr{D}_{q_{i}}^{m}U$ is the $m^{\mathrm{th}}$ order spatial
gradient, $q_{i}$ is the $i^{\mathrm{th}}$ generalized coordinate
in $d$-dimensions, and $a_{k}$ and $b_{m}$ are coefficients for
the time and spatial derivatives, respectively. For example, one-dimensional
unsteady convection-diffusion-reaction equation with a source $g\left(x,t\right)$
\citep{Kim2020} is written as
\begin{equation}
\frac{\partial C}{\partial t}=D_{0}\frac{\partial^{2}C}{\partial x^{2}}-v\frac{\partial C}{\partial x}-kC+g\left(x,t\right)
\end{equation}
where $C$ is the concentration of solutes of diffusion coefficient
$D_{0}$, convective velocity $v$, and the reaction coefficient $k$.
Mathematical parameters are determined, such as $s=1$, $a_{1}=1$,
and $a_{k\ne1}=0$ for time derivatives; $r=2$, and $d=1$ giving
$q_{1}=x$ for the spatial derivative; and $D_{0}=b_{2}$, $v=-b_{1}$,
and $k=-b_{0}$ for the physical parameters. The remainder operator
is determined, such as
\begin{equation}
R\left[U\right]=\left[\sum_{j=0}^{s-1}a_{j}D_{t}^{j}-\sum_{k=0}^{r}b_{k}\mathscr{D}_{q}^{k}\right]U
\end{equation}
and substituted into the governing equation \eqref{eq:trans-eq} to
create a perturbation series, such as 
\begin{align}
\sum_{k=0}^{\infty}\lambda^{k}\frac{\partial^{s}U_{k}}{\partial t^{s}}+\sum_{k=0}^{\infty}\lambda^{k}R\left[U_{k}\right]+\sum_{k=0}^{\infty}\lambda^{k}A_{k} & =g\label{eq:LRN}
\end{align}
Dimensionless time and concentration are defined, such as $\tau=t/t_{0}$
and $u=U/\overline{U}$, respectively, where $t_{0}$ is a representative
time scale. Eq. \eqref{eq:LRN} is rewritten as 
\begin{align}
\sum_{k=0}^{\infty}\lambda^{k}\frac{\partial^{s}u_{k}}{\partial\tau^{s}}+\sum_{k=0}^{\infty}\lambda^{k}t_{0}^{s}R\left[u_{k}\right]+\sum_{k=0}^{\infty}\lambda^{k}t_{0}^{s}\overline{U}^{k-1}A_{k}\left(\left\{ u_{k}\right\} \right) & =gt_{0}^{s}\label{eq:gov-u-lambda}
\end{align}
where $t_{0}^{s}R$ and $t_{0}^{s}\overline{U}^{k-1}A_{k}$ are unconditionally
dimensionless. Eq. \eqref{eq:gov-u-lambda} is equivalent to employing
perturbation-style $u$, such as 
\begin{equation}
u\left(\lambda\right)=u_{0}+\lambda u_{1}+\lambda^{2}u_{2}+\cdots=\sum_{n=0}^{\infty}\lambda^{n}u_{n}\label{eq:u-lambda}
\end{equation}
Without loosing generality, we let 
\begin{align}
\overline{U} & =1\\
t_{0}^{s} & =\lambda
\end{align}
to rewrite Eq. \eqref{eq:gov-u-lambda}, such as
\begin{align}
\sum_{k=0}^{\infty}\lambda^{k}\left[\frac{\partial^{s}u_{k}}{\partial\tau^{s}}+\lambda R\left[u_{k}\right]+\lambda A_{k}\left(\left\{ u_{k}\right\} \right)\right] & =g\lambda\label{eq:u-perturb}
\end{align}
which is similar to a standard perturbation series having $\lambda$
as an expansion parameter. Eq. \eqref{eq:u-perturb} specifically
gives
\begin{align}
\frac{\partial^{s}u_{0}}{\partial\tau^{s}} & -g\lambda=0
\end{align}
for $k=0$ and 
\begin{align}
\frac{\partial^{s}u_{k}}{\partial\tau^{s}} & =-\lambda R\left[u_{k}\right]-\lambda A_{k}\left(\left\{ u_{j\le k}\right\} \right)\label{eq:u-perturb-each}
\end{align}
for $k>0$.  By applying the inverse operator $L^{-1}$ from the
left on the both side of Eq. \eqref{eq:u-perturb-each}, we obtain
a new recurrence formula of Eq. \eqref{eq:ADM-recur}, such as
\begin{align}
u_{k+1} & =-\lambda R\left[L^{-1}\left(u_{k}\right)\right]-\lambda A_{k}\left(L^{-1}\left\{ u_{j\le k}\right\} \right)\label{eq:ADM-recur-perturb}
\end{align}
using the commutative relationships of 
\begin{align}
L^{-1}R\left[u_{k}\right] & =R\left[L^{-1}u_{k}\right]\label{eq:commu-R-1}\\
L^{-1}A_{k}\left[\left\{ u_{j\le k}\right\} \right] & =A_{k}\left[L^{-1}\left\{ u_{j\le k}\right\} \right]\label{eq:commu-N-1}
\end{align}
where $u_{j\le k}$ is obtained when $u_{k+1}$ is sought. In our
view, Eq. \eqref{eq:ADM-recur-perturb} resolves the Adomian's dilemma
of the order mismatch of the expansion parameter by employing nonlinear
perturbation theory with a physically dimensioned expansion parameter.
Once sufficient component functions with a large $n$ are obtained,
the truncated form of $u$ is finally obtained by letting $\lambda=1$
in Eq. \eqref{eq:u-lambda} as an analytical series solution.

\section{Concluding Remarks}

Applying Adomian's decomposition led to a recurrence formula that
efficiently provides series solutions for nonlinear governing equations,
regardless of the physical dimensions. Lyapunov's technique utilizes
a small artificial parameter, presenting a physical quantity of a
finite dimension. While intuitively proven, the Adomian decomposition
method was perceived intuitively as a special case of Lyapunov's model
instead of the perturbation theory having a closer similarity. The
present work employed the nonlinear perturbation theory with a physically
dimensioned expansion parameter and resolved the order mismatch dilemma
of the expansion parameter. To have rigorous recursion processes for
a large $n$, a new formalism is of great necessity for advanced ADM
applications, ensuring faster convergence and higher accuracy of truncated
solutions.

\bmhead{Acknowledgments}
This work has been supported in part by the U.S. National Science Foundation (grant no. 2034824).



 


\begin{thebibliography}{44}
\ifx \bisbn   \undefined \def \bisbn  #1{ISBN #1}\fi
\ifx \binits  \undefined \def \binits#1{#1}\fi
\ifx \bauthor  \undefined \def \bauthor#1{#1}\fi
\ifx \batitle  \undefined \def \batitle#1{#1}\fi
\ifx \bjtitle  \undefined \def \bjtitle#1{#1}\fi
\ifx \bvolume  \undefined \def \bvolume#1{\textbf{#1}}\fi
\ifx \byear  \undefined \def \byear#1{#1}\fi
\ifx \bissue  \undefined \def \bissue#1{#1}\fi
\ifx \bfpage  \undefined \def \bfpage#1{#1}\fi
\ifx \blpage  \undefined \def \blpage #1{#1}\fi
\ifx \burl  \undefined \def \burl#1{\textsf{#1}}\fi
\ifx \doiurl  \undefined \def \doiurl#1{\url{https://doi.org/#1}}\fi
\ifx \betal  \undefined \def \betal{\textit{et al.}}\fi
\ifx \binstitute  \undefined \def \binstitute#1{#1}\fi
\ifx \binstitutionaled  \undefined \def \binstitutionaled#1{#1}\fi
\ifx \bctitle  \undefined \def \bctitle#1{#1}\fi
\ifx \beditor  \undefined \def \beditor#1{#1}\fi
\ifx \bpublisher  \undefined \def \bpublisher#1{#1}\fi
\ifx \bbtitle  \undefined \def \bbtitle#1{#1}\fi
\ifx \bedition  \undefined \def \bedition#1{#1}\fi
\ifx \bseriesno  \undefined \def \bseriesno#1{#1}\fi
\ifx \blocation  \undefined \def \blocation#1{#1}\fi
\ifx \bsertitle  \undefined \def \bsertitle#1{#1}\fi
\ifx \bsnm \undefined \def \bsnm#1{#1}\fi
\ifx \bsuffix \undefined \def \bsuffix#1{#1}\fi
\ifx \bparticle \undefined \def \bparticle#1{#1}\fi
\ifx \barticle \undefined \def \barticle#1{#1}\fi
\bibcommenthead
\ifx \bconfdate \undefined \def \bconfdate #1{#1}\fi
\ifx \botherref \undefined \def \botherref #1{#1}\fi
\ifx \url \undefined \def \url#1{\textsf{#1}}\fi
\ifx \bchapter \undefined \def \bchapter#1{#1}\fi
\ifx \bbook \undefined \def \bbook#1{#1}\fi
\ifx \bcomment \undefined \def \bcomment#1{#1}\fi
\ifx \oauthor \undefined \def \oauthor#1{#1}\fi
\ifx \citeauthoryear \undefined \def \citeauthoryear#1{#1}\fi
\ifx \endbibitem  \undefined \def \endbibitem {}\fi
\ifx \bconflocation  \undefined \def \bconflocation#1{#1}\fi
\ifx \arxivurl  \undefined \def \arxivurl#1{\textsf{#1}}\fi
\csname PreBibitemsHook\endcsname

\bibitem[\protect\citeauthoryear{Adomian}{1963}]{Adomian1963}
\begin{barticle}
\bauthor{\bsnm{Adomian}, \binits{G.}}:
\batitle{Linear {{Stochastic Operators}}}.
\bjtitle{Reviews of Modern Physics}
\bvolume{35}(\bissue{1}),
\bfpage{185}--\blpage{207}
(\byear{1963})
\doiurl{10.1103/RevModPhys.35.185}
\end{barticle}
\endbibitem

\bibitem[\protect\citeauthoryear{Adomian}{1976}]{Adomian1976}
\begin{barticle}
\bauthor{\bsnm{Adomian}, \binits{G.}}:
\batitle{Nonlinear stochastic differential equations}.
\bjtitle{Journal of Mathematical Analysis and Applications}
\bvolume{55}(\bissue{2}),
\bfpage{441}--\blpage{452}
(\byear{1976})
\doiurl{10.1016/0022-247X(76)90174-8}
\end{barticle}
\endbibitem

\bibitem[\protect\citeauthoryear{Adomian and
  Malakian}{1980a}]{Adomian.Malakian1980}
\begin{barticle}
\bauthor{\bsnm{Adomian}, \binits{G.}},
\bauthor{\bsnm{Malakian}, \binits{K.}}:
\batitle{Operator-theoretic solution of stochastic systems}.
\bjtitle{Journal of Mathematical Analysis and Applications}
\bvolume{76}(\bissue{1}),
\bfpage{183}--\blpage{201}
(\byear{1980})
\doiurl{10.1016/0022-247X(80)90071-2}
\end{barticle}
\endbibitem

\bibitem[\protect\citeauthoryear{Adomian and
  Malakian}{1980b}]{Adomian.Malakian1980a}
\begin{barticle}
\bauthor{\bsnm{Adomian}, \binits{G.}},
\bauthor{\bsnm{Malakian}, \binits{K.}}:
\batitle{Self-correcting approximate solution by the iterative method for
  linear and nonlinear stochastic differential equations}.
\bjtitle{Journal of Mathematical Analysis and Applications}
\bvolume{76}(\bissue{2}),
\bfpage{309}--\blpage{327}
(\byear{1980})
\doiurl{10.1016/0022-247X(80)90035-9}
\end{barticle}
\endbibitem

\bibitem[\protect\citeauthoryear{Adomian}{1983}]{Adomian1983}
\begin{bbook}
\bauthor{\bsnm{Adomian}, \binits{G.}}:
\bbtitle{Stochastic Systems}.
\bsertitle{Mathematics in Science and Engineering},
vol. \bseriesno{169}.
\bpublisher{Academic press},
\blocation{New York, London, Paris}
(\byear{1983})
\end{bbook}
\endbibitem

\bibitem[\protect\citeauthoryear{Adomian et~al.}{1983}]{Adomian.Bellomo.ea1983}
\begin{barticle}
\bauthor{\bsnm{Adomian}, \binits{G.}},
\bauthor{\bsnm{Bellomo}, \binits{N.}},
\bauthor{\bsnm{Riganti}, \binits{R.}}:
\batitle{Semilinear stochastic systems: {{Analysis}} with the method of the
  stochastic {{Green}}'s function and application in mechanics}.
\bjtitle{Journal of Mathematical Analysis and Applications}
\bvolume{96}(\bissue{2}),
\bfpage{330}--\blpage{340}
(\byear{1983})
\doiurl{10.1016/0022-247X(83)90044-6}
\end{barticle}
\endbibitem

\bibitem[\protect\citeauthoryear{Adomian}{1986}]{Adomian1986}
\begin{barticle}
\bauthor{\bsnm{Adomian}, \binits{G.}}:
\batitle{A new approach to the heat equation - {{An}} application of the
  decomposition method}.
\bjtitle{Journal of Mathematical Analysis and Applications}
\bvolume{113}(\bissue{1}),
\bfpage{202}--\blpage{209}
(\byear{1986})
\doiurl{10.1016/0022-247X(86)90344-6}
\end{barticle}
\endbibitem

\bibitem[\protect\citeauthoryear{Adomian}{1990}]{Adomian1990}
\begin{barticle}
\bauthor{\bsnm{Adomian}, \binits{G.}}:
\batitle{A review of the decomposition method and some recent results for
  nonlinear equations}.
\bjtitle{Mathematical and Computer Modelling}
\bvolume{13}(\bissue{7}),
\bfpage{17}--\blpage{43}
(\byear{1990})
\doiurl{10.1016/0895-7177(90)90125-7}
\end{barticle}
\endbibitem

\bibitem[\protect\citeauthoryear{Adomian}{1991}]{Adomian1991}
\begin{barticle}
\bauthor{\bsnm{Adomian}, \binits{G.}}:
\batitle{Solving frontier problems modelled by nonlinear partial differential
  equations}.
\bjtitle{Computers \& Mathematics with Applications}
\bvolume{22}(\bissue{8}),
\bfpage{91}--\blpage{94}
(\byear{1991})
\doiurl{10.1016/0898-1221(91)90017-X}
\end{barticle}
\endbibitem

\bibitem[\protect\citeauthoryear{Adomian}{1994}]{adomianSolvingFrontierProblems1994}
\begin{bbook}
\bauthor{\bsnm{Adomian}, \binits{G.}}:
\bbtitle{Solving {{Frontier Problems}} of {{Physics}}: {{The Decomposition
  Method}}}.
\bpublisher{Springer},
\blocation{Dordrecht}
(\byear{1994}).
\doiurl{10.1007/978-94-015-8289-6}
\end{bbook}
\endbibitem

\bibitem[\protect\citeauthoryear{{Mohyud-Din}
  et~al.}{2017}]{Mohyud-DinEtAl2017}
\begin{barticle}
\bauthor{\bsnm{{Mohyud-Din}}, \binits{S.T.}},
\bauthor{\bsnm{Sikander}, \binits{W.}},
\bauthor{\bsnm{Khan}, \binits{U.}},
\bauthor{\bsnm{Ahmed}, \binits{N.}}:
\batitle{Optimal variational iteration method using {{Adomian}}'s polynomials
  for physical problems on finite and semi-infinite intervals}.
\bjtitle{The European Physical Journal Plus}
\bvolume{132}(\bissue{5}),
\bfpage{236}
(\byear{2017})
\doiurl{10.1140/epjp/i2017-11506-9}
\end{barticle}
\endbibitem

\bibitem[\protect\citeauthoryear{Hamrelaine et~al.}{2022}]{HamrelaineEtAl2022}
\begin{barticle}
\bauthor{\bsnm{Hamrelaine}, \binits{S.}},
\bauthor{\bsnm{Kezzar}, \binits{M.}},
\bauthor{\bsnm{Sari}, \binits{M.R.}},
\bauthor{\bsnm{Eid}, \binits{M.R.}}:
\batitle{Analytical investigation of hydromagnetic ferro-nanofluid flowing via
  rotating convergent/divergent channels}.
\bjtitle{The European Physical Journal Plus}
\bvolume{137}(\bissue{11}),
\bfpage{1291}
(\byear{2022})
\doiurl{10.1140/epjp/s13360-022-03480-2}
\end{barticle}
\endbibitem

\bibitem[\protect\citeauthoryear{Panda et~al.}{2022}]{PandaEtAl2022}
\begin{barticle}
\bauthor{\bsnm{Panda}, \binits{A.}},
\bauthor{\bsnm{Santra}, \binits{S.}},
\bauthor{\bsnm{Mohapatra}, \binits{J.}}:
\batitle{Adomian decomposition and homotopy perturbation method for the
  solution of time fractional partial integro-differential equations}.
\bjtitle{Journal of Applied Mathematics and Computing}
\bvolume{68}(\bissue{3}),
\bfpage{2065}--\blpage{2082}
(\byear{2022})
\doiurl{10.1007/s12190-021-01613-x}
\end{barticle}
\endbibitem

\bibitem[\protect\citeauthoryear{Singh and Wazwaz}{2022}]{SinghWazwaz2022}
\begin{barticle}
\bauthor{\bsnm{Singh}, \binits{R.}},
\bauthor{\bsnm{Wazwaz}, \binits{A.-M.}}:
\batitle{Analytical approximations of three-point generalized
  {{Thomas}}--{{Fermi}} and {{Lane}}--{{Emden}}--{{Fowler}} type equations}.
\bjtitle{The European Physical Journal Plus}
\bvolume{137}(\bissue{1}),
\bfpage{63}
(\byear{2022})
\doiurl{10.1140/epjp/s13360-021-02301-2}
\end{barticle}
\endbibitem

\bibitem[\protect\citeauthoryear{Naghdi}{2023}]{Naghdi2023}
\begin{barticle}
\bauthor{\bsnm{Naghdi}, \binits{M.}}:
\batitle{Solutions for scalar equations in {{AdS}}$_4$ with {{Adomian}} method
  and boundary {{CFT}}$_3$ duals}.
\bjtitle{The European Physical Journal Plus}
\bvolume{138}(\bissue{3}),
\bfpage{300}
(\byear{2023})
\doiurl{10.1140/epjp/s13360-023-03905-6}
\end{barticle}
\endbibitem

\bibitem[\protect\citeauthoryear{Abdelrazec and
  Pelinovsky}{2011}]{Abdelrazec.Pelinovsky2011}
\begin{barticle}
\bauthor{\bsnm{Abdelrazec}, \binits{A.}},
\bauthor{\bsnm{Pelinovsky}, \binits{D.}}:
\batitle{{Convergence of the Adomian decomposition method for initial-value
  problems}}.
\bjtitle{Numerical Methods for Partial Differential Equations}
\bvolume{27}(\bissue{4}),
\bfpage{749}--\blpage{766}
(\byear{2011})
\doiurl{10.1002/num.20549}
\end{barticle}
\endbibitem

\bibitem[\protect\citeauthoryear{Zhang et~al.}{2018}]{Zhang.Zou.ea2018}
\begin{barticle}
\bauthor{\bsnm{Zhang}, \binits{X.}},
\bauthor{\bsnm{Zou}, \binits{L.}},
\bauthor{\bsnm{Liang}, \binits{S.}},
\bauthor{\bsnm{Liu}, \binits{C.}}:
\batitle{A novel analytic approximation method with a convergence acceleration
  parameter for solving nonlinear problems}.
\bjtitle{Communications in Nonlinear Science and Numerical Simulation}
\bvolume{56},
\bfpage{354}--\blpage{364}
(\byear{2018})
\doiurl{10.1016/j.cnsns.2017.08.025}
\end{barticle}
\endbibitem

\bibitem[\protect\citeauthoryear{Adomian}{1995}]{Adomian1995}
\begin{barticle}
\bauthor{\bsnm{Adomian}, \binits{G.}}:
\batitle{Solving the mathematical models of neurosciences and medicine}.
\bjtitle{Mathematics and Computers in Simulation}
\bvolume{40}(\bissue{1}),
\bfpage{107}--\blpage{114}
(\byear{1995})
\doiurl{10.1016/0378-4754(95)00021-8}
\end{barticle}
\endbibitem

\bibitem[\protect\citeauthoryear{Adomian and Rach}{1996}]{Adomian.Rach1996}
\begin{barticle}
\bauthor{\bsnm{Adomian}, \binits{G.}},
\bauthor{\bsnm{Rach}, \binits{R.}}:
\batitle{Modified {{Adomian Polynomials}}}.
\bjtitle{Mathematical and Computer Modelling}
\bvolume{24}(\bissue{11}),
\bfpage{39}--\blpage{46}
(\byear{1996})
\doiurl{10.1016/S0895-7177(96)00171-9}
\end{barticle}
\endbibitem

\bibitem[\protect\citeauthoryear{{Al-Shareef}
  et~al.}{2016}]{Al-Shareef.AlQarni.ea2016}
\begin{barticle}
\bauthor{\bsnm{{Al-Shareef}}, \binits{A.}},
\bauthor{\bsnm{Al~Qarni}, \binits{A.A.}},
\bauthor{\bsnm{{Al-Mohalbadi}}, \binits{S.}},
\bauthor{\bsnm{Bakodah}, \binits{H.O.}}:
\batitle{Soliton {{Solutions}} and {{Numerical Treatment}} of the {{Nonlinear
  Schrodinger}}'s {{Equation Using Modified Adomian Decomposition Method}}}.
\bjtitle{Journal of Applied Mathematics and Physics}
\bvolume{04}(\bissue{12}),
\bfpage{2215}--\blpage{2232}
(\byear{2016})
\doiurl{10.4236/jamp.2016.412215}
\end{barticle}
\endbibitem

\bibitem[\protect\citeauthoryear{Kumar and {Umesh}}{2022}]{Kumar.Umesh2022}
\begin{barticle}
\bauthor{\bsnm{Kumar}, \binits{M.}},
\bauthor{\bsnm{{Umesh}}}:
\batitle{Recent {{Development}} of {{Adomian Decomposition Method}} for
  {{Ordinary}} and {{Partial Differential Equations}}}.
\bjtitle{International Journal of Applied and Computational Mathematics}
\bvolume{8}(\bissue{2}),
\bfpage{81}
(\byear{2022})
\doiurl{10.1007/s40819-022-01285-6}
\end{barticle}
\endbibitem

\bibitem[\protect\citeauthoryear{Biazar et~al.}{2010}]{BiazarEtAl2010}
\begin{barticle}
\bauthor{\bsnm{Biazar}, \binits{J.}},
\bauthor{\bsnm{Gholami~Porshokuhi}, \binits{M.}},
\bauthor{\bsnm{Ghanbari}, \binits{B.}}:
\batitle{Extracting a general iterative method from an {{Adomian}}
  decomposition method and comparing it to the variational iteration method}.
\bjtitle{Computers \& Mathematics with Applications}
\bvolume{59}(\bissue{2}),
\bfpage{622}--\blpage{628}
(\byear{2010})
\doiurl{10.1016/j.camwa.2009.11.001}
\end{barticle}
\endbibitem

\bibitem[\protect\citeauthoryear{{Daftardar-Gejji} and
  Bhalekar}{2008}]{Daftardar-GejjiBhalekar2008}
\begin{barticle}
\bauthor{\bsnm{{Daftardar-Gejji}}, \binits{V.}},
\bauthor{\bsnm{Bhalekar}, \binits{S.}}:
\batitle{Solving multi-term linear and non-linear diffusion--wave equations of
  fractional order by {{Adomian}} decomposition method}.
\bjtitle{Applied Mathematics and Computation}
\bvolume{202}(\bissue{1}),
\bfpage{113}--\blpage{120}
(\byear{2008})
\doiurl{10.1016/j.amc.2008.01.027}
\end{barticle}
\endbibitem

\bibitem[\protect\citeauthoryear{Sadeghinia and
  Kumar}{2021}]{SadeghiniaKumar2021}
\begin{bchapter}
\bauthor{\bsnm{Sadeghinia}, \binits{A.}},
\bauthor{\bsnm{Kumar}, \binits{P.}}:
\bctitle{One {{Solution}} of {{Multi-term Fractional Differential Equations}}
  by {{Adomian Decomposition Method}}: {{Scientific Explanation}}}.
In: \beditor{\bsnm{Wang}, \binits{D.X.}} (ed.)
\bbtitle{Current {{Topics}} on {{Mathematics}} and {{Computer Science Vol}}.
  6},
pp. \bfpage{120}--\blpage{130}.
\bpublisher{Book Publisher International}, \blocation{London, UK}
(\byear{2021}).
\doiurl{10.9734/bpi/ctmcs/v6/11542D}
\end{bchapter}
\endbibitem

\bibitem[\protect\citeauthoryear{Zhang and Liang}{2015}]{Zhang.Liang2015}
\begin{barticle}
\bauthor{\bsnm{Zhang}, \binits{X.}},
\bauthor{\bsnm{Liang}, \binits{S.}}:
\batitle{Adomian decomposition method is a special case of {{Lyapunov}}'s
  artificial small parameter method}.
\bjtitle{Applied Mathematics Letters}
\bvolume{48},
\bfpage{177}--\blpage{179}
(\byear{2015})
\doiurl{10.1016/j.aml.2015.04.011}
\end{barticle}
\endbibitem

\bibitem[\protect\citeauthoryear{Adomian and Rach}{1983}]{Adomian.Rach1983}
\begin{barticle}
\bauthor{\bsnm{Adomian}, \binits{G.}},
\bauthor{\bsnm{Rach}, \binits{R.}}:
\batitle{Inversion of nonlinear stochastic operators}.
\bjtitle{Journal of Mathematical Analysis and Applications}
\bvolume{91}(\bissue{1}),
\bfpage{39}--\blpage{46}
(\byear{1983})
\doiurl{10.1016/0022-247X(83)90090-2}
\end{barticle}
\endbibitem

\bibitem[\protect\citeauthoryear{Rach}{1984}]{Rach1984}
\begin{barticle}
\bauthor{\bsnm{Rach}, \binits{R.}}:
\batitle{A convenient computational form for the {{Adomian}} polynomials}.
\bjtitle{Journal of Mathematical Analysis and Applications}
\bvolume{102}(\bissue{2}),
\bfpage{415}--\blpage{419}
(\byear{1984})
\doiurl{10.1016/0022-247X(84)90181-1}
\end{barticle}
\endbibitem

\bibitem[\protect\citeauthoryear{Rach}{2008}]{Rach2008}
\begin{barticle}
\bauthor{\bsnm{Rach}, \binits{R.C.}}:
\batitle{A new definition of the {{Adomian}} polynomials}.
\bjtitle{Kybernetes}
\bvolume{37}(\bissue{7}),
\bfpage{910}--\blpage{955}
(\byear{2008})
\doiurl{10.1108/03684920810884342}
\end{barticle}
\endbibitem

\bibitem[\protect\citeauthoryear{Bell}{1986}]{Bell1986}
\begin{bbook}
\bauthor{\bsnm{Bell}, \binits{E.T.}}:
\bbtitle{Men of Mathematics},
\bedition{1st touchstone ed} edn.
\bsertitle{A {{Touchstone}} Book}.
\bpublisher{Simon \& Schuster},
\blocation{New York}
(\byear{1986})
\end{bbook}
\endbibitem

\bibitem[\protect\citeauthoryear{Lagrange}{1772}]{Lagrange1772}
\begin{bbook}
\bauthor{\bsnm{Lagrange}, \binits{J.-L.}}:
\bbtitle{Essai Sur Le Probl{\`e}me Des Trois Corps ({{Essay}} on the {{Essay}}
  on the Three-Body Problem)}
vol. \bseriesno{6},
(\byear{1772})
\end{bbook}
\endbibitem

\bibitem[\protect\citeauthoryear{Laplace}{1798}]{Laplace1798}
\begin{bbook}
\bauthor{\bsnm{Laplace}, \binits{P.S.}}:
\bbtitle{{Trait{\'e} de M{\'e}canique C{\'e}leste}}.
\bpublisher{De L'Imprimerie de Crapelet : Chez J.B.M. Duprat}, \blocation{Paris, France}
(\byear{1798})
\end{bbook}
\endbibitem

\bibitem[\protect\citeauthoryear{Poincar{\'e}}{1890}]{Poincare1890}
\begin{bbook}
\bauthor{\bsnm{Poincar{\'e}}, \binits{H.}}:
\bbtitle{The Three-Body Problem and the Equations of Dynamics:
  {{Poincar{\'e}}}'s Foundational Work on Dynamical Systems Theory}.
\bsertitle{Astrophysics and Space Science Library},
vol. \bseriesno{443}.
\bpublisher{Springer},
\blocation{Cham, Switzerland}
(\byear{1890})
\end{bbook}
\endbibitem

\bibitem[\protect\citeauthoryear{Schr{\"o}dinger}{1926a}]{Schrodinger1926}
\begin{barticle}
\bauthor{\bsnm{Schr{\"o}dinger}, \binits{E.}}:
\batitle{Quantisierung als {{Eigenwertproblem}}}.
\bjtitle{Annalen der Physik}
\bvolume{384}(\bissue{4}),
\bfpage{361}--\blpage{376}
(\byear{1926})
\doiurl{10.1002/andp.19263840404}
\end{barticle}
\endbibitem

\bibitem[\protect\citeauthoryear{Schr{\"o}dinger}{1926b}]{Schrodinger1926a}
\begin{barticle}
\bauthor{\bsnm{Schr{\"o}dinger}, \binits{E.}}:
\batitle{An {{Undulatory Theory}} of the {{Mechanics}} of {{Atoms}} and
  {{Molecules}}}.
\bjtitle{Physical Review}
\bvolume{28}(\bissue{6}),
\bfpage{1049}--\blpage{1070}
(\byear{1926})
\doiurl{10.1103/PhysRev.28.1049}
\end{barticle}
\endbibitem

\bibitem[\protect\citeauthoryear{Dirac}{1955}]{Dirac1955}
\begin{barticle}
\bauthor{\bsnm{Dirac}, \binits{P.A.M.}}:
\batitle{Note on the {{Use}} of {{Non-Orthogonal Wave Functions}} in
  {{Perturbation Calculations}}}.
\bjtitle{Canadian Journal of Physics}
\bvolume{33}(\bissue{12}),
\bfpage{709}--\blpage{712}
(\byear{1955})
\doiurl{10.1139/p55-087}
\end{barticle}
\endbibitem

\bibitem[\protect\citeauthoryear{Feynman}{1949}]{Feynman1949}
\begin{barticle}
\bauthor{\bsnm{Feynman}, \binits{R.P.}}:
\batitle{Space-{{Time Approach}} to {{Quantum Electrodynamics}}}.
\bjtitle{Physical Review}
\bvolume{76}(\bissue{6}),
\bfpage{769}--\blpage{789}
(\byear{1949})
\doiurl{10.1103/PhysRev.76.769}
\end{barticle}
\endbibitem

\bibitem[\protect\citeauthoryear{Wazwaz}{2000}]{Wazwaz2000}
\begin{barticle}
\bauthor{\bsnm{Wazwaz}, \binits{A.-M.}}:
\batitle{A new algorithm for calculating adomian polynomials for nonlinear
  operators}.
\bjtitle{Applied Mathematics and Computation}
\bvolume{111}(\bissue{1}),
\bfpage{33}--\blpage{51}
(\byear{2000})
\doiurl{10.1016/S0096-3003(99)00063-6}
\end{barticle}
\endbibitem

\bibitem[\protect\citeauthoryear{Fatoorehchi and
  Abolghasemi}{2016}]{Fatoorehchi.Abolghasemi2016}
\begin{barticle}
\bauthor{\bsnm{Fatoorehchi}, \binits{H.}},
\bauthor{\bsnm{Abolghasemi}, \binits{H.}}:
\batitle{Series solution of nonlinear differential equations by a novel
  extension of the {{Laplace}} transform method}.
\bjtitle{International Journal of Computer Mathematics}
\bvolume{93}(\bissue{8}),
\bfpage{1299}--\blpage{1319}
(\byear{2016})
\doiurl{10.1080/00207160.2015.1045421}
\end{barticle}
\endbibitem

\bibitem[\protect\citeauthoryear{Kataria and
  Vellaisamy}{2016}]{katariaSimpleParametrizationMethods2016}
\begin{barticle}
\bauthor{\bsnm{Kataria}, \binits{K.K.}},
\bauthor{\bsnm{Vellaisamy}, \binits{P.}}:
\batitle{Simple parametrization methods for generating {{Adomian}}
  polynomials}.
\bjtitle{Applicable Analysis and Discrete Mathematics}
\bvolume{10}(\bissue{1}),
\bfpage{168}--\blpage{185}
(\byear{2016})
\doiurl{10.2298/AADM160123001K}
\end{barticle}
\endbibitem

\bibitem[\protect\citeauthoryear{Fatoorehchi and
  Abolghasemi}{2013}]{fatoorehchiImprovingDifferentialTransform2013}
\begin{barticle}
\bauthor{\bsnm{Fatoorehchi}, \binits{H.}},
\bauthor{\bsnm{Abolghasemi}, \binits{H.}}:
\batitle{Improving the differential transform method: {{A}} novel technique to
  obtain the differential transforms of nonlinearities by the {{Adomian}}
  polynomials}.
\bjtitle{Applied Mathematical Modelling}
\bvolume{37}(\bissue{8}),
\bfpage{6008}--\blpage{6017}
(\byear{2013})
\doiurl{10.1016/j.apm.2012.12.007}
\end{barticle}
\endbibitem

\bibitem[\protect\citeauthoryear{Kao and Jiang}{2005}]{Kao.Jiang2005}
\begin{barticle}
\bauthor{\bsnm{Kao}, \binits{Y.-M.}},
\bauthor{\bsnm{Jiang}, \binits{T.F.}}:
\batitle{Adomian's decomposition method for eigenvalue problems}.
\bjtitle{Physical Review E}
\bvolume{71}(\bissue{3}),
\bfpage{036702}
(\byear{2005})
\doiurl{10.1103/PhysRevE.71.036702}
\end{barticle}
\endbibitem

\bibitem[\protect\citeauthoryear{Benettin
  et~al.}{1980}]{Benettin.Galgani.ea1980}
\begin{barticle}
\bauthor{\bsnm{Benettin}, \binits{G.}},
\bauthor{\bsnm{Galgani}, \binits{L.}},
\bauthor{\bsnm{Giorgilli}, \binits{A.}},
\bauthor{\bsnm{Strelcyn}, \binits{J.-M.}}:
\batitle{Lyapunov {{Characteristic Exponents}} for smooth dynamical systems and
  for hamiltonian systems; {{A}} method for computing all of them. {{Part}} 2:
  {{Numerical}} application}.
\bjtitle{Meccanica}
\bvolume{15}(\bissue{1}),
\bfpage{21}--\blpage{30}
(\byear{1980})
\doiurl{10.1007/BF02128237}
\end{barticle}
\endbibitem

\bibitem[\protect\citeauthoryear{Wolf et~al.}{1985}]{Wolf.Swift.ea1985}
\begin{barticle}
\bauthor{\bsnm{Wolf}, \binits{A.}},
\bauthor{\bsnm{Swift}, \binits{J.B.}},
\bauthor{\bsnm{Swinney}, \binits{H.L.}},
\bauthor{\bsnm{Vastano}, \binits{J.A.}}:
\batitle{Determining {{Lyapunov}} exponents from a time series}.
\bjtitle{Physica D: Nonlinear Phenomena}
\bvolume{16}(\bissue{3}),
\bfpage{285}--\blpage{317}
(\byear{1985})
\doiurl{10.1016/0167-2789(85)90011-9}
\end{barticle}
\endbibitem

\bibitem[\protect\citeauthoryear{Kim}{2020}]{Kim2020}
\begin{barticle}
\bauthor{\bsnm{Kim}, \binits{A.S.}}:
\batitle{Complete analytic solutions for convection-diffusion-reaction-source
  equations without using an inverse {{Laplace}} transform}.
\bjtitle{Scientific Reports}
\bvolume{10}(\bissue{1}),
\bfpage{8040}
(\byear{2020})
\doiurl{10.1038/s41598-020-63982-w}
\end{barticle}
\endbibitem

\end{thebibliography}

\begin{appendices}

\section{Understanding $A_n$ coefficient}

In the ADM, the true solution $u$ is defined as 
\begin{equation}
u=\sum_{n=0}^{\infty}u_{n}
\end{equation}
and the nonlinear term is assumed to be an analytic function of $u$,
i.e., $N\left[u\right]=f\left(u\right)$, such as 
\begin{equation}
f\left(u\right)=\sum_{n=0}^{\infty}A_{n}\left(u_{0},\cdots,u_{n}\right)
\end{equation}
where $u_{0}$ is the trivial solution in the absence of $R$ and
$N$, and $A_{n}$ is known as the $n^{\mathrm{th}}$-order Adomian
polynomial. The $A_{n}$ is given as
\begin{equation}
A_{n}=\sum_{\nu=1}^{n}f_{0}^{\left(\nu\right)}C\left(\nu,n\right)
\end{equation}
where
\begin{equation}
f_{0}^{\left(\nu\right)}=\left[\frac{\mathrm{d}^{\nu}f\left(u\right)}{\mathrm{d}u^{\nu}}\right]_{u=u_{0}}
\end{equation}
 is the $\nu^{\mathrm{th}}$-derivative of $f\left(u\right)$ with
respect to $u$ at $u=u_{0}$, and 
\begin{equation}
C\left(\nu,n\right)=\sum_{\nu_{i}}\prod_{k_{i}}^{\nu}\frac{1}{k_{i}!}u_{\nu_{i}}^{k_{i}}
\end{equation}
include product sums of $\nu$ components of $u$ whose subscripts
sum to $n$, divided by the factorial of the number of repeated subscripts.
For $n^{\mathrm{th}}$ order, the following conditions should be satisfied,
such as
\begin{equation}
\sum_{i=1}k_{i}=\nu\label{eq:nu-constr}
\end{equation}
and
\begin{equation}
\sum_{i=1}^{\nu_{i}}k_{i}\nu_{i}=n\label{eq:n-constr}
\end{equation}
where $k_{i}$ is the number of repetitions of $u_{\nu_{i}}$ with
$0\le i\le n$.  For example, $A_{4}$ consists of products of four
$u$'s, such as 
\begin{equation}
\frac{u_{i}^{a}}{a!}\frac{u_{j}^{b}}{b!}\frac{u_{k}^{c}}{c!}\frac{u_{l}^{d}}{d!}
\end{equation}
where $a$ is the number of repetition of $u_{i}$, and similarly
for $b$, $c$, and $d$, satisfying conditions of 
\begin{align}
a+b+c+d & =\nu\\
a\cdot i+b\cdot j+c\cdot k+d\cdot l & =n
\end{align}
and
\begin{equation}
1\le i<j<k<l\le n
\end{equation}
For $n=10$, $C\left(4,10\right)$ can be written explicitly, such
as 
\begin{align}
C\left(4,10\right) 
& =\left(\frac{1}{2!}u_{2}^{2}\right)\left(\frac{1}{2!}u_{3}^{2}\right)
+\left(\frac{1}{1!}u_{1}^{1}\right)\left(\frac{1}{3!}u_{3}^{3}\right)
\nonumber \\&
+\left(\frac{1}{1!}u_{1}^{1}\right)\left(\frac{1}{1!}u_{2}^{1}\right)\left(\frac{1}{1!}u_{3}^{1}\right)\left(\frac{1}{1!}u_{4}^{1}\right)
+\left(\frac{1}{3!}u_{2}^{3}\right)\left(\frac{1}{1!}u_{4}^{1}\right)\
\nonumber \\ & 
+\left(\frac{1}{2!}u_{1}^{2}\right)\left(\frac{1}{2!}u_{4}^{2}\right)+\left(\frac{1}{1!}u_{1}^{1}\right)\left(\frac{1}{2!}u_{2}^{2}\right)\left(\frac{1}{1!}u_{5}^{1}\right)\nonumber \\
 & +\left(\frac{1}{2!}u_{1}^{2}\right)\left(\frac{1}{1!}u_{3}^{1}\right)\left(\frac{1}{1!}u_{5}^{1}\right)+\left(\frac{1}{2!}u_{1}^{2}\right)\left(\frac{1}{1!}u_{2}^{1}\right)\left(\frac{1}{1!}u_{6}^{1}\right)
\nonumber \\&
+\left(\frac{1}{3!}u_{1}^{3}\right)\left(\frac{1}{1!}u_{7}^{1}\right)
\end{align}
having each term as a product of four components, satisfying
\begin{align}
\sum_{i=1}k_{i} & =a+b+c+d=4\\
\sum_{i=1}^{\nu_{i}}k_{i}\nu_{i} & =a\cdot i+b\cdot j+c\cdot k+d\cdot l=10
\end{align}
 For example, the first term of $C\left(4,10\right)$ can be written
as
\begin{equation}
\left(\frac{1}{2!}u_{2}^{2}\right)\left(\frac{1}{2!}u_{3}^{2}\right)=\frac{1}{2!}\frac{1}{2!}u_{2}^{1}u_{2}^{1}u_{3}^{1}u_{3}^{1}
\end{equation}
where the sum of superscripts is $1+1+1+1=4$, the product sum of
subscripts and superscripts is $1\cdot2+1\cdot2+1\cdot3+1\cdot3=10$,
and each of $u_{2}$ and $u_{3}$ are repeated twice providing $1/\left(2!2!\right)$.
All other terms follow the same rules for the subscript and superscript
sums.  When $n$ is large even if $\nu$ is small (e.g., $\nu=4$),
finding all the possible combinations satisfying Eqs. \eqref{eq:nu-constr}
and \eqref{eq:n-constr} is a challenging task. The highest order
of $A_{n}$ reported in the literature is 10 from one of Adomian's
initial work \citep{adomianSolvingFrontierProblems1994}.

\end{appendices}
\end{document}